\RequirePackage{lineno}
\documentclass[twocolumn,prc,aps,showpacs,floatfix,amsmath,amssymb]{revtex4}

\usepackage{graphicx}   
\usepackage{placeins}
\usepackage{subcaption}
\usepackage{array}

\usepackage{hyperref}
\hypersetup{
    colorlinks,
    citecolor=blue,
    filecolor=blue,
    linkcolor=blue,
    urlcolor=blue
}
\newcommand{\sqrtsNN}{\mbox{$\sqrt{\mathrm{s}_{_{\mathrm{NN}}}}$}}

\newcommand{\ppt}{$p_{\rm T}$}

\newcommand{\muB}{$\mu_B$}
\def \auau  {Au+Au}

\begin{document}

\begin{center}
{\bf Effect of hadronic cascade time on freeze-out properties of Identified Hadrons in {\auau} Collisions at {\sqrtsNN} = 7.7 - 39 GeV from AMPT Model}
\end{center}

\author{
M.~U.~Ashraf$^{1}${\footnote{usman.ashraf@cern.ch; }},
Junaid~Tariq$^{2}$ {\footnote{junaid.tariqhep@gmail.com; }} and
A.~M.~Khan$^{3}$ {\footnote{ahsan.mehmood.khan@cern.ch; }}
}

\affiliation{
$^1$ Centre for Cosmology, Particle Physics and Phenomenology (CP3), Université Catholique de Louvain, B-1348 Louvain-la-Neuve, Belgium\\
$^2$ Department of Physics, Quaid-i-Azam University, Islamabad 44000, Pakistan\\
$^3$Key Laboratory of Quark \& Lepton Physics (MOE) and Institute of Particle Physics,
Central China Normal University, Wuhan 430079, China\\
}

\date{\today}

\begin{abstract}
  
We report the transverse momentum {\ppt} spectra of identified hadrons ($\pi^\pm$, $K^\pm$ and $p(\bar p)$) in {\auau} collisions at {\sqrtsNN} = 7.7 - 39 GeV from A Multi Phase Transport Model with string melting effect (AMPT-SM). During this study, a new set of parameters are explored to study the effect of hadronic cascade by varying hadronic cascade time $t_{max}$ = 30 $f$m/$c$ and 0.4 $f$m/$c$. No significant effect of this change is observed in the {\ppt} spectra of light hadrons and the AMPT-SM model reasonably reproduces the experimental data. To investigate the kinetic freeze-out properties the blast wave fit is performed to the {\ppt} spectra and it is found that the blast wave model describes the AMPT-SM simulations well. We additionally observe that the kinetic freeze-out temperature ($T_{kin}$) increases from central to peripheral collisions, which is consistent with the argument of short-lived fireball in peripheral collisions. Whereas the transverse flow velocity, $<\beta_T>$ shows a decreasing trend from central to peripheral collisions indicating a more rapid expansion in the central collisions. Both, $T_{kin}$ and $<\beta_T>$ show a weak dependence on the collision energy at most energies. We also observe a strong  anti-correlation between $T_{kin}$ and $<\beta_T>$. The extracted freeze-out parameters from the AMPT-SM simulations agree with the experimental data as opposed to earlier studies that reported some discrepancies. Whereas, no significant effect is found on the freeze-out parameters by varying the $t_{max}$. We also report the {\ppt} spectra of light hadrons and their freeze-out parameters by AMPT-SM simulations at {\sqrtsNN} = 14.5 GeV, where no experimental data is available for comparison. Overall, the set of parameters used in this study well describes the experimental data at BES energies.     

\end{abstract}


\maketitle

\section{Introduction}
\label{Introduction}
Theoretical work utilizing quantum chromodynamics (QCD) strongly suggests of the existence of a deconfined state of quarks and gluons at high temperatures and/or high baryon densities.  Current theoretical models predict the possible existence of a QCD critical point, which may occur at the edge of the boundary of the first order phase transition at low temperatures and high baryon number chemical potentials ({\muB}) regions~\cite{1, 2, 3, 4, 5}. 
 At the QCD critical point, the transition behavior of ordinary nuclear matter into an amalgamation of free quarks and gluons, known as a quark-gluon plasma (QGP) changes instantaneously. 
  Beyond this critical point, in the regions of relatively low {\muB} and high $T$'s, QCD additionally predicts a continuous and smooth crossover from the hadron gas phase to the QGP~\cite{6}. Description and explanation of the space-time evolution of this deconfined matter is inherently complex if only one or even many of the available theoretical models are used.  This complexity arises due to different degrees of freedoms involved under various space-time coordinates.

 Ultra-relativistic heavy-ion collisions are employed as the main tool to study the deconfined state of nuclear matter.
  QCD phase diagram can be mapped at various temperatures and baryon chemical  potentials to study the different phases of matter. 
  To accomplish this, the relativistic heavy ion collider (RHIC) at Brookhaven National Laboratory (BNL) undertook the Beam Energy Scan phase 1 (BES-I) program from 2010 to 2017 and reported on the {\auau} collisions at {\sqrtsNN} = 7.7 - 39 GeV~\cite{7, 8, 9, 10, 11, 12}. 
 The main objectives of the BES program include, mapping the QCD phase diagram, locating the QCD critical point and finding the boundary region between the two phases. 


The bulk properties of a collision system are essential tools that provide information on how the system evolves over time. These properties assist in our understanding of the expansion of the fireball resulting from the heavy-ion interactions. In heavy-ion interactions, two possible scenarios of freeze-out are observed, the chemical freeze-out and the kinetic freeze-out. During the chemical freeze-out, the inelastic collisions between the hadrons stop, which means that there will be no new bound states produced after this stage.
During the chemical freeze-out, multiple thermodynamical models~\cite{13, 14, 15, 16} provide the means for the extraction of {\muB} and chemical freeze-out temperature ($T_{ch}$). The chemical freeze-out stage is followed by the kinetic freeze-out stage. The time of the hadronic phase between chemical and kinetic freeze-out is proportional to a parameter called hadron cascade time ($\tau_{HC}$). The study of kinetic freeze-out stage is complex, however, it is vital to undertake because multiple literature report various freeze-out scenarios~\cite{17, 18, 19, 20, 21}. Additionally, the kinetic freeze-out temperature ($T_{kin}$) is directly related to QGP temperature, which depends on the density or the number of participating nucleons. Hence, the study of the increase in $T_{kin}$ from central to peripheral collisions make it vital to our understanding of freeze-out~\cite{ 22, Lokesh}. Earlier study~\cite{48} suggests a clear discrepancy in the freeze-out parameters between AMPT simulations and experimental data. 
 
 In this paper we compare the experimental data collected at the Solenoidal Tracker at RHIC (STAR) experiment to the data obtained using the A Multi-Phase Transport model (AMPT) simulations. Here the kinetic freeze-out temperature ($T_{kin}$) and the radial flow velocity ($\beta_T$) parameters are extracted by fitting the AMPT-SM simulation. The extraction is accomplished by fitting the Blast-wave function to the simulated {\ppt} spectra of identified charged particles, $\pi^\pm$, $K^\pm$ and $p(\overline{p}$). The freeze-out parameters are also studied as a function of centrality and collision energy. Further, we used two values, 30 $f$m/$c$ and 0.4 $f$m/$c$, of the hadron cascade time, $\tau_{HC}$, which is referred to as $t_{max}$, in the AMPT-SM model. 

The paper is organized as follows: Section~\ref{sec2} gives the brief description of the AMPT model, followed by a results and discussion in section~\ref{sec3} and finally a conclusion provided in section~\ref{Conclusion}.

 \section{A multi-phase transport (AMPT) model}
 \label{sec2}
In this section, a short description of the AMPT model and its parameters are discussed. The AMPT model was developed to study the dynamics of relativistic heavy-ion collisions~\cite{24}. Further, this model has been extensively used to study the particle properties at various energies and for multiple colliding systems. Currently, two versions of the AMPT model, namely the default AMPT and the AMPT with string melting (AMPT-SM) version are in use. The default version of the AMPT was first released around April 2004, whereas the AMPT-SM version was introduced later~\cite{25, 26}. AMPT-SM version, also called hybrid transport model, was developed with four main components: the initial conditions, partonic interactions, hadronization, and hadronic interactions~\cite{24}. The initial conditions are based on the Heavy Ion Jet Interaction Generator (HIJING) model~\cite{27}, which includes the initial spatial and momentum distributions of minijet partons and the soft string excitations. When the momentum transfer in the production of hard minijet partons is greater than the threshold ($p_0 = 2$ GeV/$c$) the perturbative processes play an important role. While soft strings are produced when the momentum transfer is less than the threshold value. 

The scatterings among partons are modeled by Zhang’s parton cascade (ZPC)~\cite{28}. This currently includes only two-body scatterings with cross sections obtained from the perturbative quantum chromodynamics (pQCD) theory with Debye screening mass in the partonic matter. The scattering cross-section is mathematically given by:

\begin{equation}
    \sigma_p \approx \frac{9 \pi \alpha^2_s}{2(t-\mu^2)^2}
    \label{eq1}
\end{equation}

where, $\sigma_p$ is the parton-parton scattering, $t$ is the Mandelstam variable for four momentum transfer, $\alpha_s$ is the strong coupling constant and $\mu$ is the Debye screening mass in partonic matter.

Once the interaction among the partons stop, in the default AMPT version, the partons recombine with their parent strings. This recombination results in the production of hadrons utilizing the Lund string fragmentation model~\cite{29, 30}. 
Whereas, in the AMPT-SM version, all flavors of quarks and antiquarks ($q\overline{q}$) take part in the ZPC, and the hadronization takes place due to the quark coalescence model.  The quark coalescence model is responsible for the coalescence of the nearest partons to form hadrons. Additionally, in the AMPT-SM version more partons are produced per unit volume and the coalescence of quarks enhances the elliptic flow of hadrons. Therefore, the AMPT-SM model is able to better describe the large elliptic flow with small parton cross sections 
at RHIC energies~\cite{25, 26}. 
The last component of AMPT is hadronic interaction; the hadronic re-scattering process is described by a hadronic cascade, and it is based on A Relativistic Transport (ART) model~\cite{31}. The ART model describes the dynamics of hadronic matter including the meson-meson, meson-baryon, baryon-baryon, elastic, and inelastic scatterings~\cite{31}.  Final observables from the AMPT model are obtained after the hadronic interactions cease at a certain cutoff time ($t_{cut}$). The cutoff time ($t_{cut}$) is when the observables (final results) under study are considered stable and do not change significantly due to further interactions. Hence $t_{cut}$ provides the time limit for the hadronic interaction.

In this study, we use the AMPT-SM version with the following set of parameters, strong coupling constant ($\alpha_s$) = 0.33 and the parton screening mass ($\mu$) = 3.20 $f$m$^{-1}$, which gives the value of $\sigma_p$ = 1.5 $mb$ using equation~\ref{eq1}. We also use an improved quark coalescence method for the current study~\cite{32} and ART model for the dynamics of hadronic matter, however, we did not use the mean field, which describes the potentials of hadrons, in the ART model to carry out our analysis~\cite{33}. In the improved quark coalescence method, the relative probability of a quark forming a baryon rather than a meson can be controlled by a new coalescence parameter $r_{BM}$, which is set to be 0.61 for our study. The $r_{BM}$ parameter well describes the proton yield $dN/dy$ at mid-rapidity in central {\auau} collisions at $\sqrtsNN$ = 200 GeV as well as central $Pb$-$Pb$ collisions at $\sqrtsNN$ = 2.76 TeV~\cite{32}. Baryons can be produced either in pairs ($B\bar B$) or in combination with mesons ($BM\bar B$ ). This production of baryons is explained by a method called the popcorn method and it is controlled in the AMPT-SM model by a popcorn parameter called PARJ(5). In this study, to control the relative percentage of the $B\bar B$ and $BM\bar B$ channels the value of PARJ(5) is changed from default value of 1.0 to 0.0.  

\section{Analysis Methodology}
\label{sec3}
This section focuses on the results obtained during this study and then discusses the impact of these results. We start with the results obtained for the transverse momentum ({\ppt}) of the identified charged hadrons and followed by the results obtained for their kinetic properties, temperature ($T_{kin}$) and the transverse flow velocity ($<\beta_T>$).

\subsection{Transverse Momentum ({\ppt}) spectra}

In this study, using an improved version of AMPT-SM model and $\sigma_p$ = 1.5 $mb$, we obtain the freeze-out properties of the identified hadrons in {\auau} collisions. Motivated by the RHIC beam energy scan (BES-I) program~\cite{14}, $2 \times 10^4$ events were generated with the improved AMPT-SM at the following {\sqrtsNN} = 7.7, 11.5, 14.5, 19.6, 27 and 39 GeV. The transverse momentum ({\ppt}) spectra within the rapidity $|y|< 0.1$ of identified hadrons are measured and the effect of hadronic interaction on the {\ppt} as well as the freeze-out properties are studied by varying the hadronic cascade time ($t_{max}$).

\begin{figure*}[!ht]{\label{fig1}}
\centering
\begin{subfigure}{0.8\textwidth}
\includegraphics[width=\textwidth]{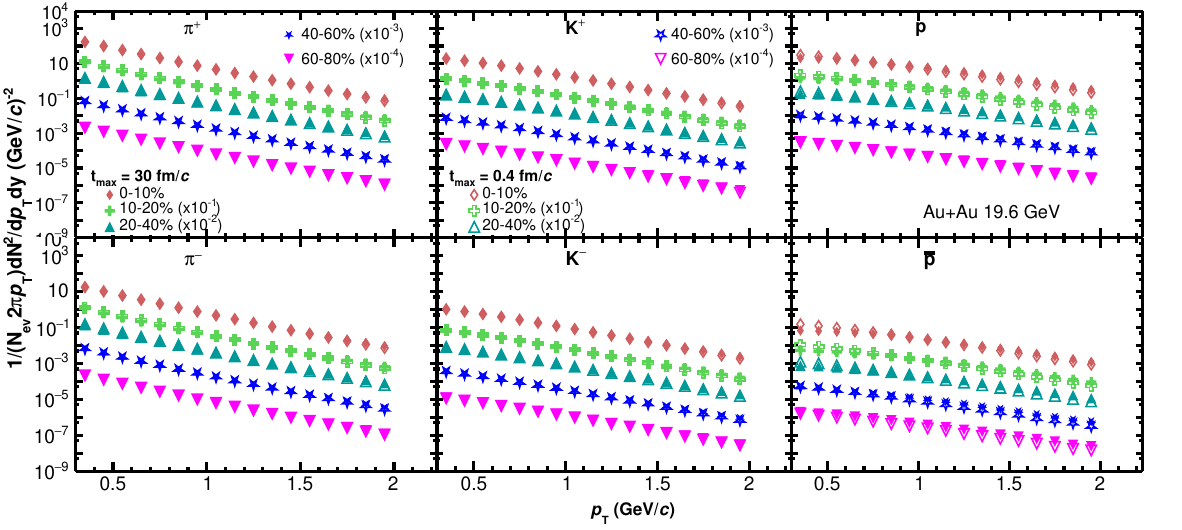}
\vspace{-1.0\baselineskip}
\end{subfigure}
\begin{subfigure}{0.8\textwidth}
\includegraphics[width=\textwidth]{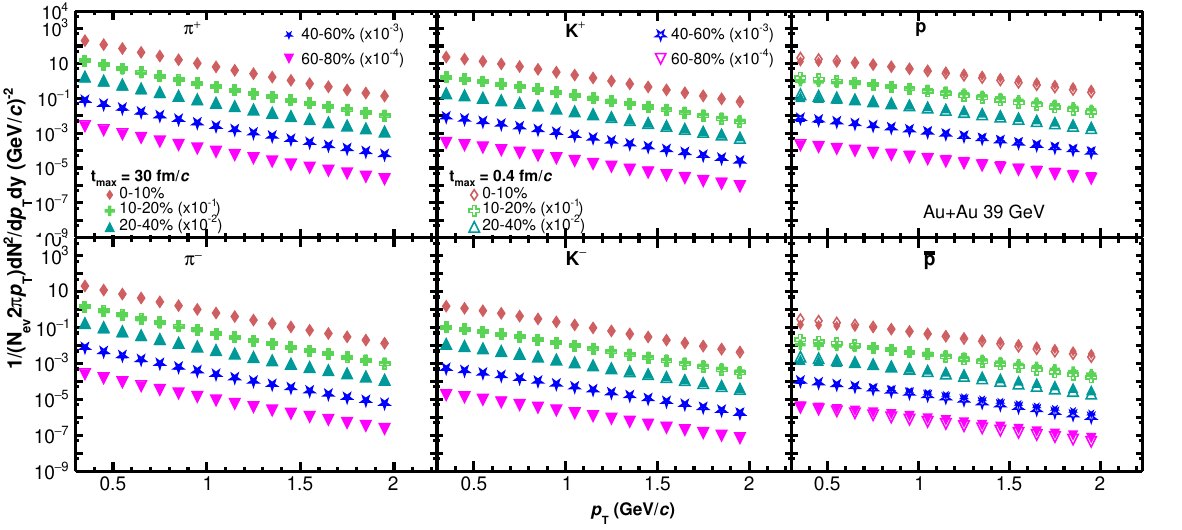}
\vspace{-1.0\baselineskip}
\end{subfigure}
\caption{The transverse momentum ({\ppt}) spectra of $\pi^\pm$, $K^\pm$, $p$ and $\bar p$ at midrapidity $|y|< 0.1$ in {\auau} collisions at {\sqrtsNN} = 19.6 and 39 GeV from the AMPT-SM model at different centralities. The {\ppt} spectra is scaled by a factor of $10$ for better visualization. Solid markers represents the {\ppt} spectra obtained from $t_{max}$ = 30 $f$m/$c$, while open markers shows the {\ppt} spectra from $t_{max}$ = 0.4 $f$m/$c$. } 
\label{fig1}
\end{figure*}

Figure~\ref{fig1} shows the transverse momentum ({\ppt}) spectra in {\auau} collisions of $\pi^{\pm}$, $K^\pm$, $p$ and $\bar p$ at midrapidity $|y|< 0.1$ at {\sqrtsNN} = 19.6 GeV (upper panel) and 39 GeV (lower panel) respectively from the improved AMPT-SM model at given centralities, from most central (0-10\%) to increasingly peripheral values (10-20\%, 20-40\%, 40-6\% and 60-80\%). In the fig.~\ref{fig1}, solid symbols represent {\ppt} spectra with $t_{max}$ = 30 $f$m/$c$ while open symbols represent $t_{max}$ = 0.4 $f$m/$c$. This figure shows that the invariant yield of all the identified hadrons decreases from central to peripheral collisions. The effect of hadronic cascade is studied by changing $t_{max}$ from 30 $f$m/$c$ to 0.4 $f$m/$c$.  However, comparing the results for two values of $t_{max}$, 30 $f$m/$c$ and 0.4 $f$m/$c$, shows no significant difference for the {\ppt} spectra. It is experimentally found that the spectral shape of {\ppt} for colliding systems is exponential and for heavier particles such as the proton the slope is flatter than the slope obtained for lighter particles such as pions. This difference in the spectral shape is because of the radial flow effects~\cite{14, 36, 37}.  We also observe a similar pattern for the {\ppt} spectra in our data as shown in fig.~\ref{fig1}.


\subsection{Comparison with Experimental Results}

 Figure~\ref{fig2} compares the {\ppt} spectra of $\pi^+$, $K^+$ and $p$ at midrapidity ($|y|< 0.1$) in the most central (0-10\%) {\auau} collisions at {\sqrtsNN} = 14.5 GeV (upper panel) and 39 GeV (lower panel) respectively from improved AMPT-SM simulations to that of STAR data~\cite{14}. The results presented here are for $t_{max}$ = 30 $f$m/$c$ and $t_{max}$ = 0.4 $f$m/$c$. It is clear from fig.~\ref{fig2} that the invariant yield of all the identified hadrons decrease with increasing {\ppt}. When comparing the inverse slopes ({\ppt}/invariant yield) of the three hadrons under study, $\pi^{+}$, $K^{+}$, $p$, we observe that they follow the order  $p > K > \pi$, which means that the proton yield changes less than kaon and pion yields when {\ppt} is varied by the same amount. Similar behaviour is observed at other energies, i.e., 7.7, 11.5, 19.6 and 27 GeV. The negatively charged particles, $\pi^{-}$, $K^{-}$, $\overline{p}$, not presented in fig.~\ref{fig2} also show similar trends at all energies which can be seen in fig.~\ref{fig1}. 
 
 It is clear from fig.~\ref{fig2} that our simulation results well describes the data from the STAR Experiment~\cite{10, 14} for $\pi^+$, $K^+$ at {\sqrtsNN} = 14.5 and 39 GeV. From the lower panel of fig.~\ref{fig2}, we see that in the Model/Data ratio the {\ppt} spectra of $p$ is over-estimated particularly at low {\ppt} bins ($p_\mathrm{T} < 1$ GeV/$c$) by the improved AMPT-SM model with the set of parameters discussed in sec.~\ref{sec2}. However, at $p_\mathrm{T} > 1$ GeV/$c$, the {\ppt} spectra of proton is well described by the AMPT-SM model. Further, in the course of this study we also compared the results at other energies {\sqrtsNN} = 7.7 - 39 GeV and we observe similar trends.  
 

\begin{figure*}[!ht]{\label{fig2}}
\begin{subfigure}{0.8\textwidth}
\includegraphics[width=\textwidth]{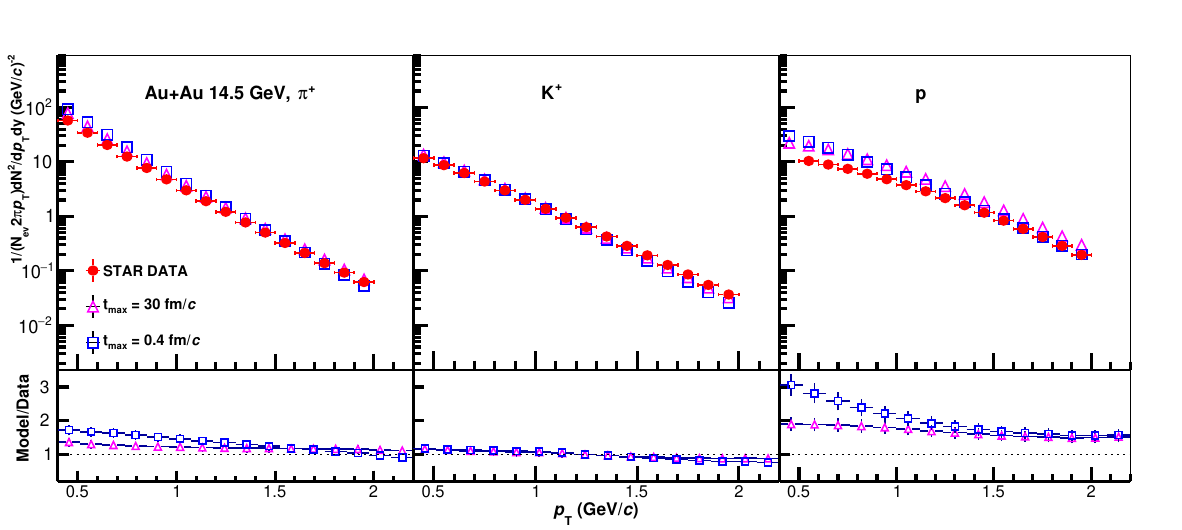}
\vspace{-1.2\baselineskip}
\label{fig2a}
\end{subfigure}
\begin{subfigure}{0.8\textwidth}
\includegraphics[width=\textwidth]{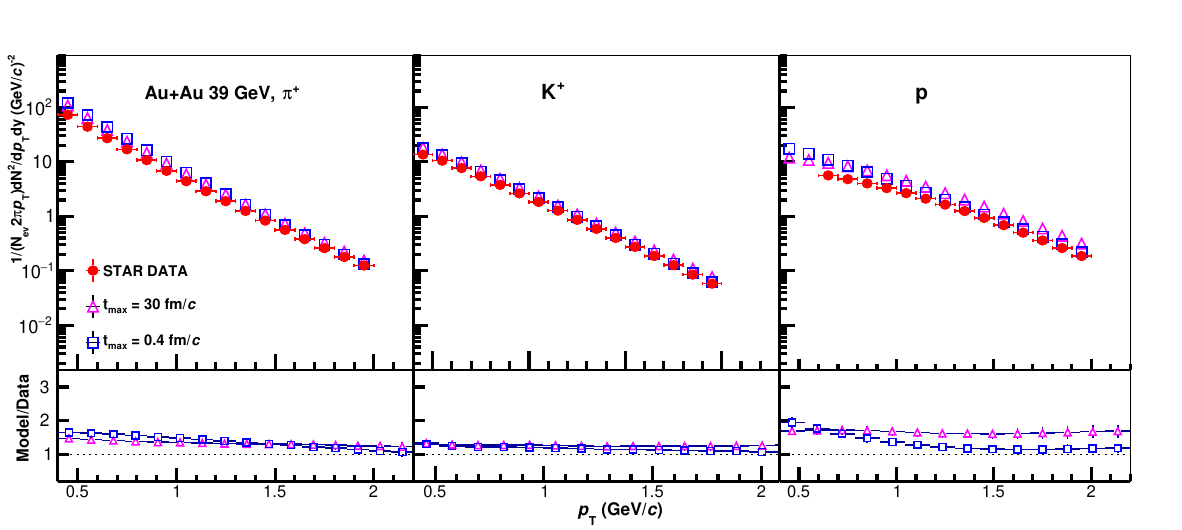}
\vspace{-1.2\baselineskip}
\label{fig2b}
\end{subfigure}
\caption{The transverse momentum ({\ppt}) spectra at midrapidity ($|y| < 0.1$) of $\pi^+$, $K^+$ and $p$ for central (0-10\%) {\auau} collisions at {\sqrtsNN} = 14.5 GeV from AMPT-SM model with $t_{max}$ = 30 $f$m/$c$ and $t_{max}$ = 0.4 $f$m/$c$. Solid markers represent experimental data from STAR~\cite{14, 38}. The bottom panels show the model-to-data ratios. } 
\label{fig2}
\end{figure*}

\subsection{Kinetic Properties}

The bulk properties of a medium can be systematically studied by measuring the {\ppt} spectra of the hadrons. In this study, we focused on the kinetic freeze-out properties extracted from the {\ppt} spectra.  As discussed above, the kinetic freeze-out stage occurs when the elastic collisions stop and spectra of the particles produced become fixed. Important parameters used to study properties of the system at this stage are the temperature ($T_{kin}$) and the transverse flow velocity ($<\beta_T>$), where $T_{kin}$ gives the temperature of the initial system and $<\beta_T>$ describes the expansion of the system in the transverse direction. The kinetic freeze-out parameters are extracted by fitting the {\ppt} spectra with a Blast Wave (BW) model~\cite{35, 39, 41} a hydrodynamics inspired model. The Blast Wave model assumes that particles are locally thermalized with a common $T_{kin}$ moving with a common transverse flow velocity $<\beta_{T}>$~\cite{35, 41}.

\begin{figure*}[!ht]{\label{fig3}}
\centering
\begin{subfigure}{0.3\textwidth}
\includegraphics[width=\textwidth]{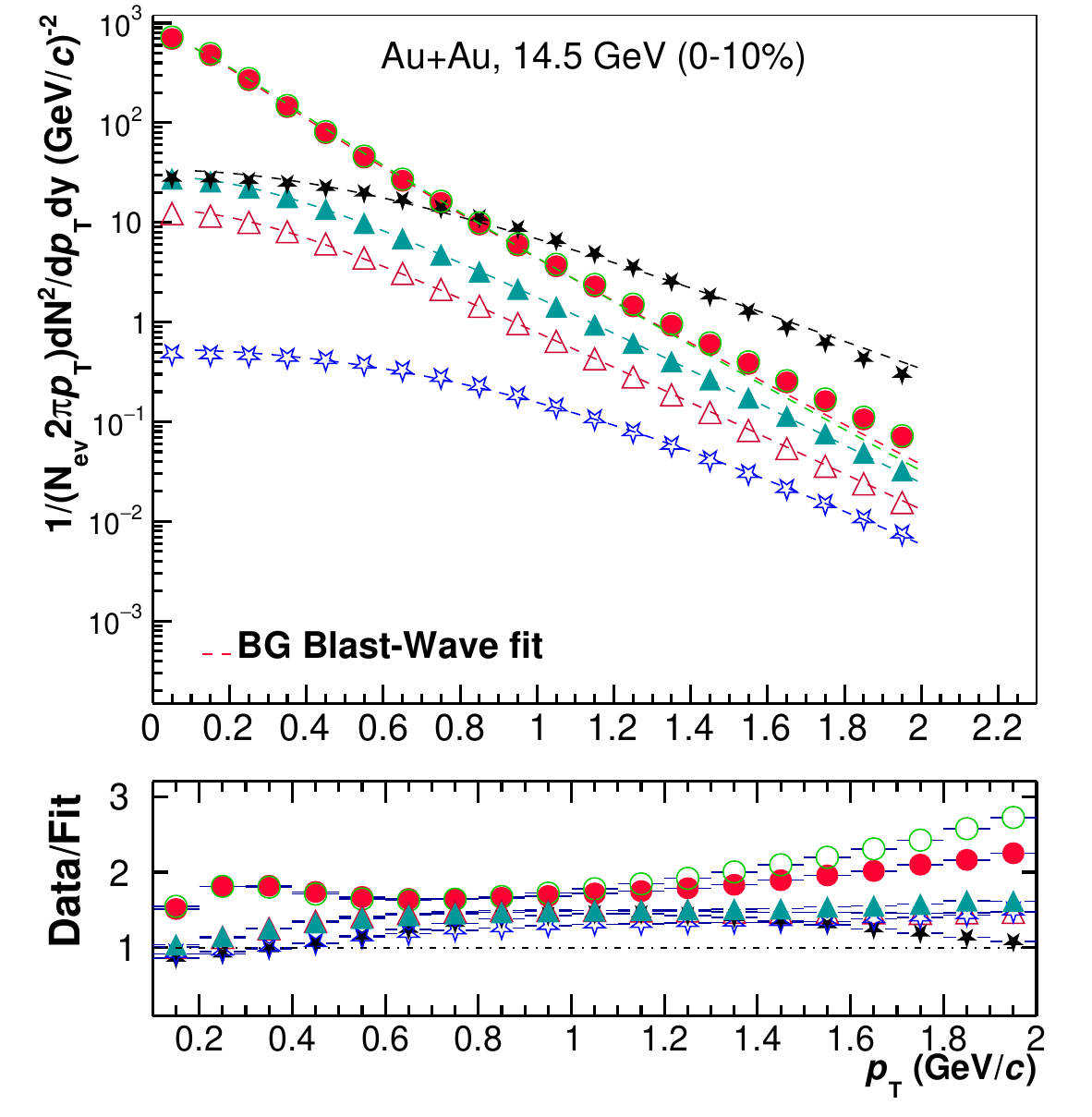}
\vspace{-1.5\baselineskip}
\caption{}
\label{fig3a}
\end{subfigure}
\begin{subfigure}{0.3\textwidth}
\includegraphics[width=\textwidth]{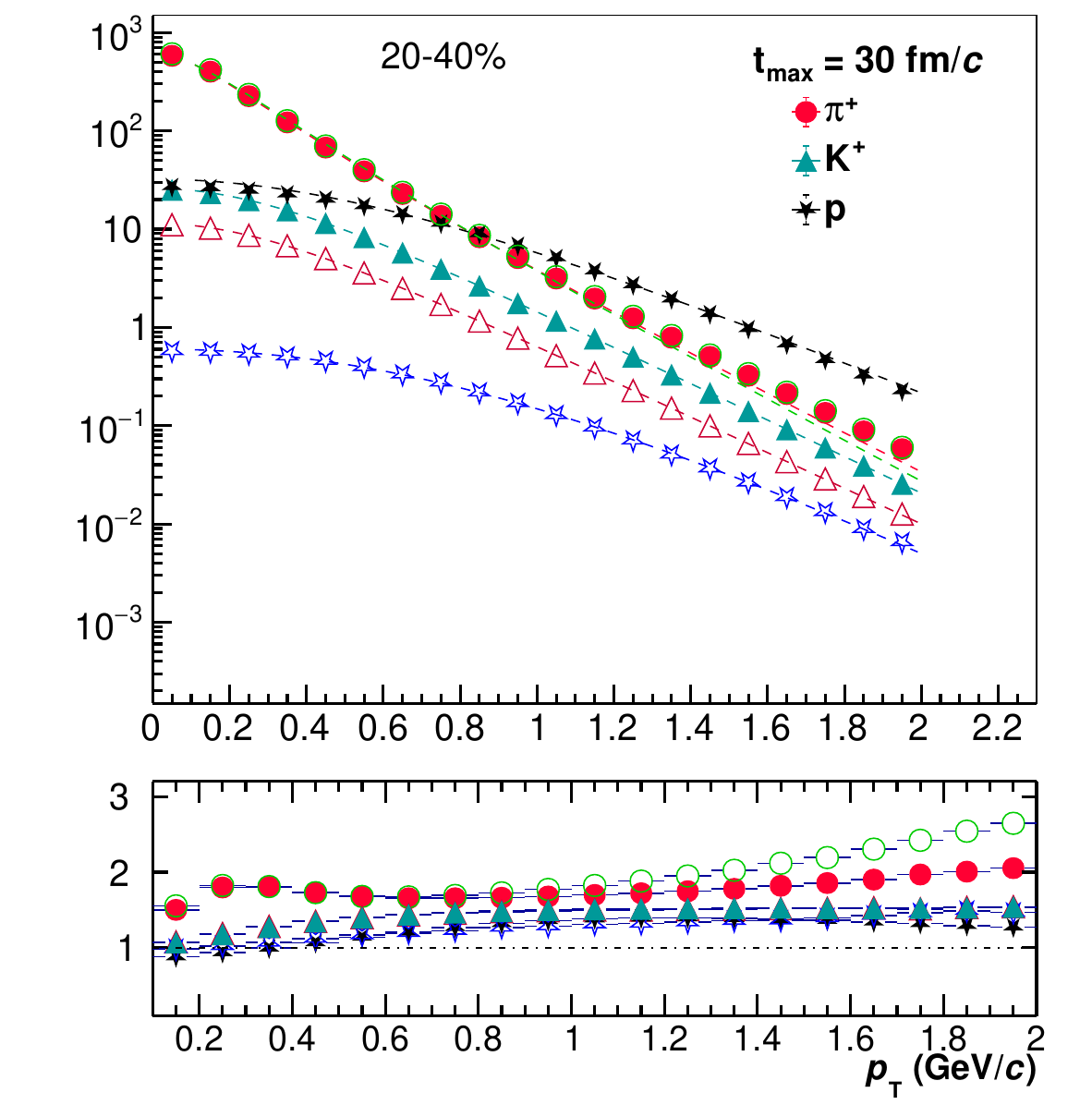}
\vspace{-1.5\baselineskip}
\caption{}
\label{fig3b}
\end{subfigure}
\begin{subfigure}{0.3\textwidth}
\includegraphics[width=\textwidth]{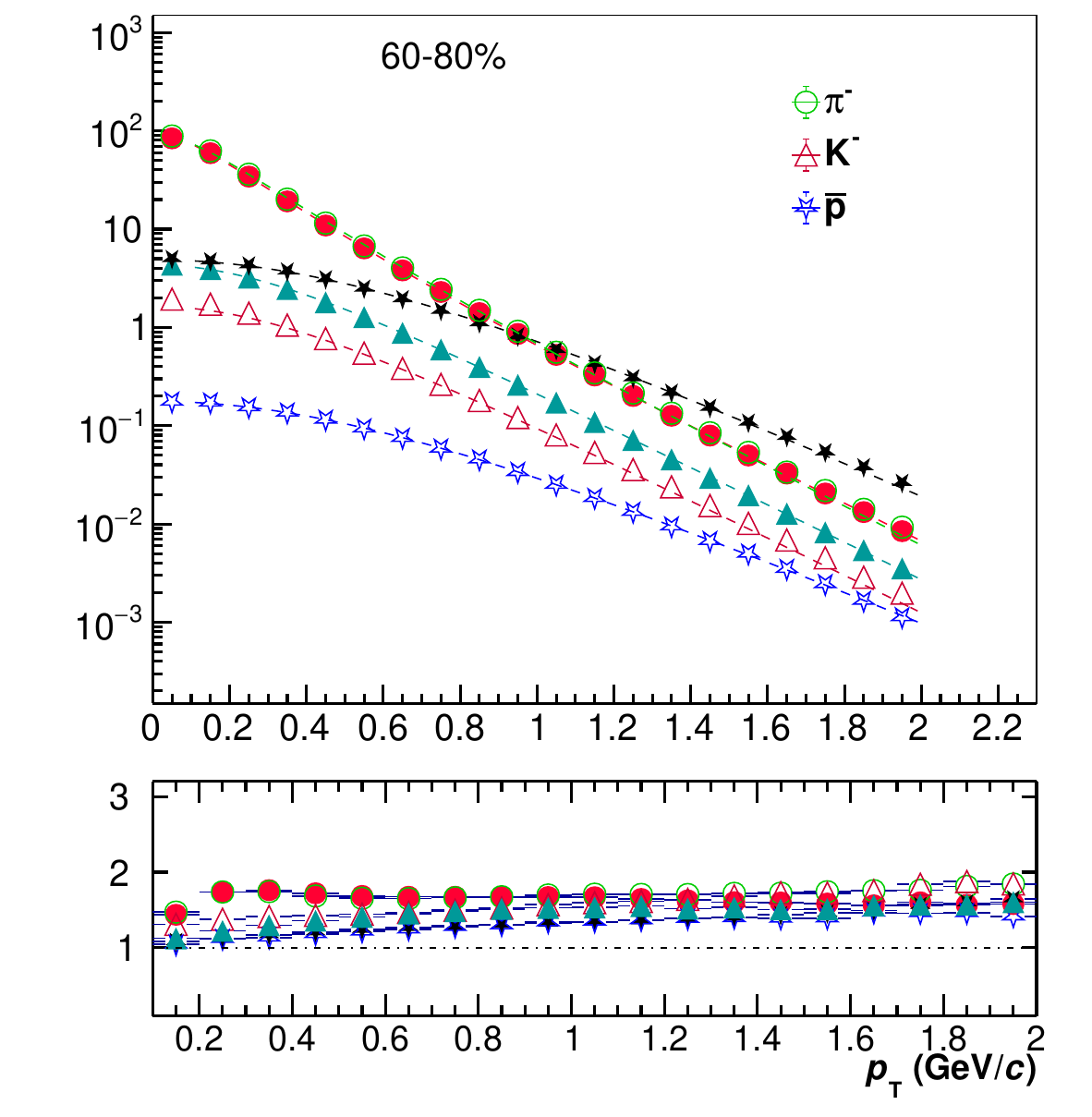}
\vspace{-1.5\baselineskip}
\caption{}
\label{fig3c}
\end{subfigure}
\begin{subfigure}{0.3\textwidth}
\includegraphics[width=\textwidth]{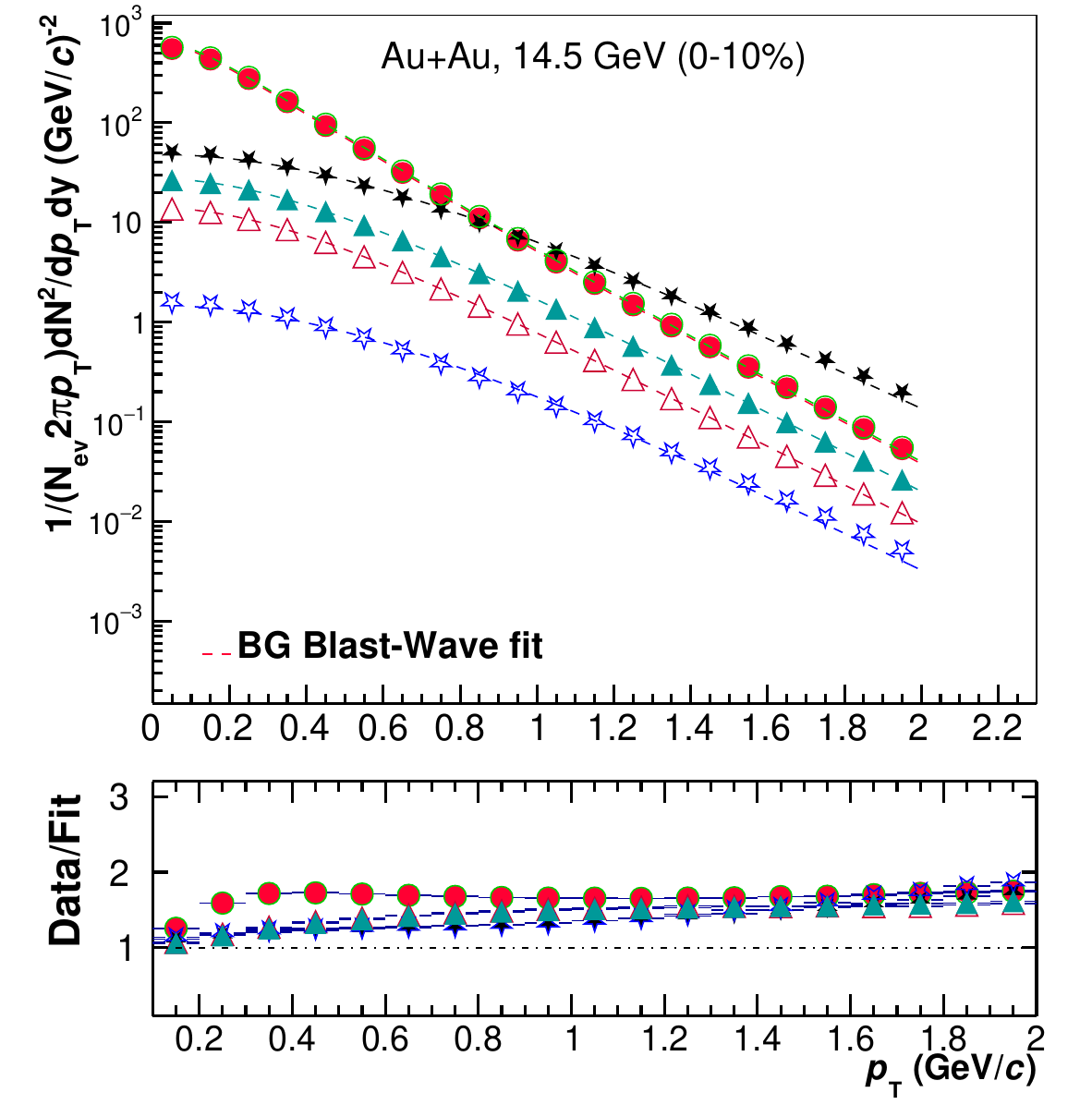}
\vspace{-1.5\baselineskip}
\caption{}
\label{fig3d}
\end{subfigure}
\begin{subfigure}{0.3\textwidth}
\includegraphics[width=\textwidth]{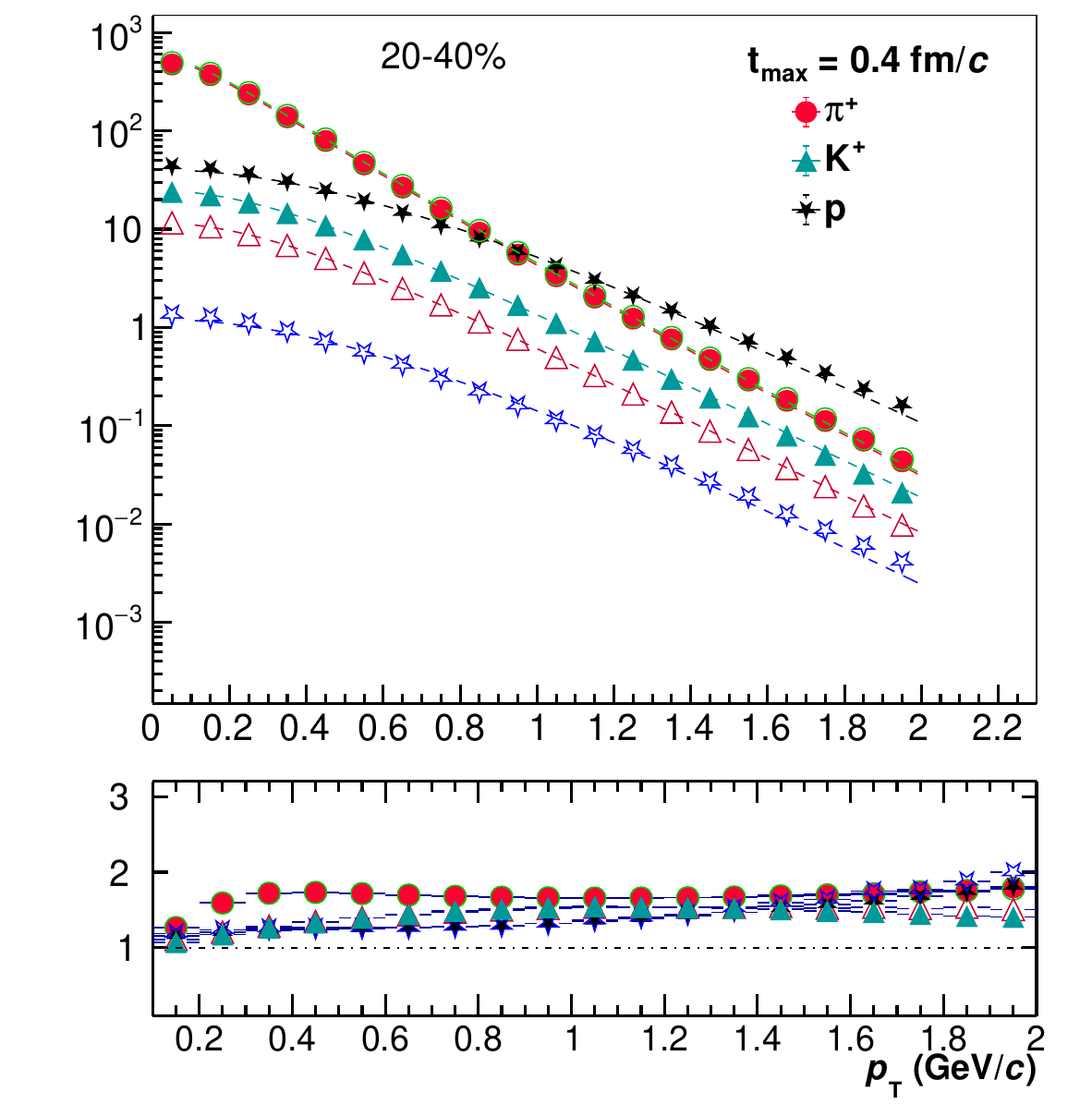}
\vspace{-1.5\baselineskip}
\caption{}
\label{fig3e}
\end{subfigure}
\begin{subfigure}{0.3\textwidth}
\includegraphics[width=\textwidth]{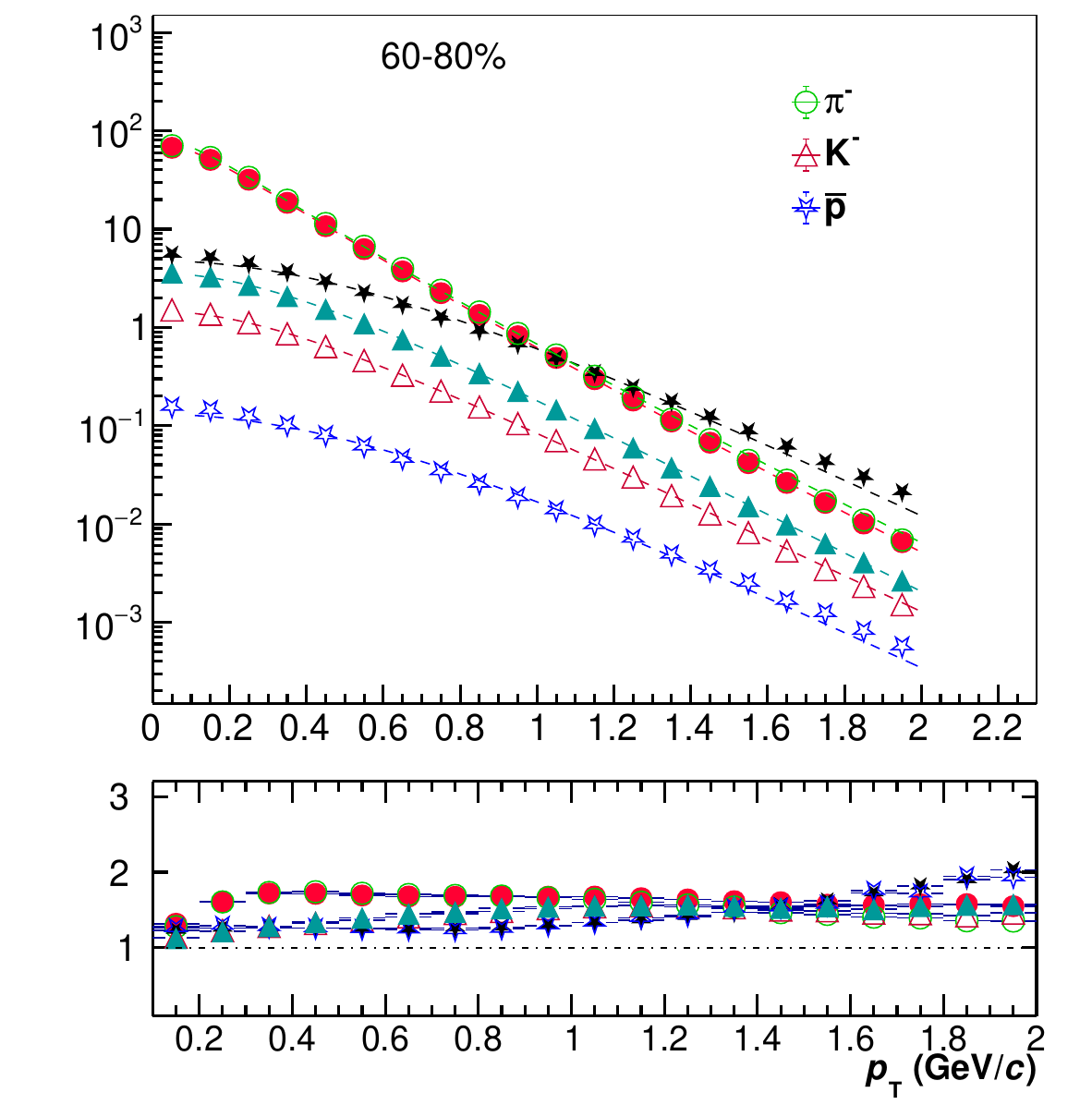}
\vspace{-1.5\baselineskip}
\caption{}
\label{fig3f}
\end{subfigure}
\caption{Blast-wave fits to hadron {\ppt} spectra in {\auau} collisions at {\sqrtsNN} = 14.5 GeV from AMPT-SM model at various centrality classes. Different symbols represent different centrality. Upper three panels (a)-(c) presents the {\ppt} spectra for $t_{max}$ = 30 $f$m/$c$ and the lower three panels (d)-(f) presents the {\ppt} spectra for $t_{max}$ = 0.4 $f$m/$c$. Solid lines presents the results for Blast-Wave fit to the {\ppt} spectra. The data-to-fit ratio is shown at the bottom of each panel.  } 
\label{fig3}
\end{figure*}
Assuming a radially boosted thermal source, with a kinetic freeze-out temperature ($T_{kin}$) and a transverse flow velocity ($\beta_T$), the {\ppt} distribution of the particles is mathematically given by:

\begin{equation}
   \frac{dN}{p_Tdp_T} \propto 
    \int_{0}^R r dr m_{T} I_{0}(\frac{p_T\sinh\rho(r)}{T_{kin}})\times K_1 (\frac{m_T\cosh\rho(r)}{T_{kin}})
\end{equation}

where $m_T = \sqrt{p_{T}^2 + m^2}$ is the transverse mass of the hadron species, $\rho \equiv tanh^{-1} \beta$, and $I_{0}$ and $K_1$ are the modified Bessel functions.

Figure~\ref{fig3} shows fit of the blast-wave function to the identified hadrons {\ppt} spectra from the AMPT-SM model in {\auau} collisions at {\sqrtsNN} = 14.5 GeV for $t_{max}$ = 30 $f$m/$c$ and 0.4 $f$m/$c$ at centrality 0-10\%, 20-40\% and 60-80\%. The data-to-fit ratio is also shown in the bottom panel. $T_{kin}$ and $<\beta_T>$ are the fit parameters and $n$ is fixed to 1.0 for the current study. $\pi^\pm$ spectra is heavily influenced by the resonance decays at low {\ppt}, hence the {\ppt} spectra of $\pi^\pm$ are only fitted for $p_\mathrm{T} > 0.5$ GeV/$c$. 

The Blast Wave is a model motivated from hydrodynamics and its fit results are sensitive to the {\ppt} ranges used for fitting~\cite{43}. The low {\ppt} part of the spectra is better described by this model than the high {\ppt} region where hard processes dominate~\cite{42}.  For the current study, we use the same value of low {\ppt} as previously reported by ALICE and STAR experiments~\cite{35, 43} and we observe that the {\ppt} spectra is well described by the Blast-wave model. It seems that the deviation of BW fit to AMPT-SM data for $t_{max}$ = 30 $f$m/$c$ is relatively large in 0-10\% and 20-40\% centrality for $p_\mathrm{T} > 1.4$ GeV/$c$ in case of $\pi^+$ and $K^+$. However, at 60-80\% centrality this deviation decreases and gives smaller value of $\chi^2/NDF$. On the other hand, for $t_{max}$ = 0.4 $f$m/$c$, the BW model well describes the AMPT-SM data for all hadron species at all centrality classes and there is a constant deviation of the BW model from the AMPT-SM data for all centralities. Overall, the BW model well describes the AMPT-SM data.

\begin{figure*}[!ht]{\label{fig4}} 
    \centering
    \includegraphics[width=\textwidth]{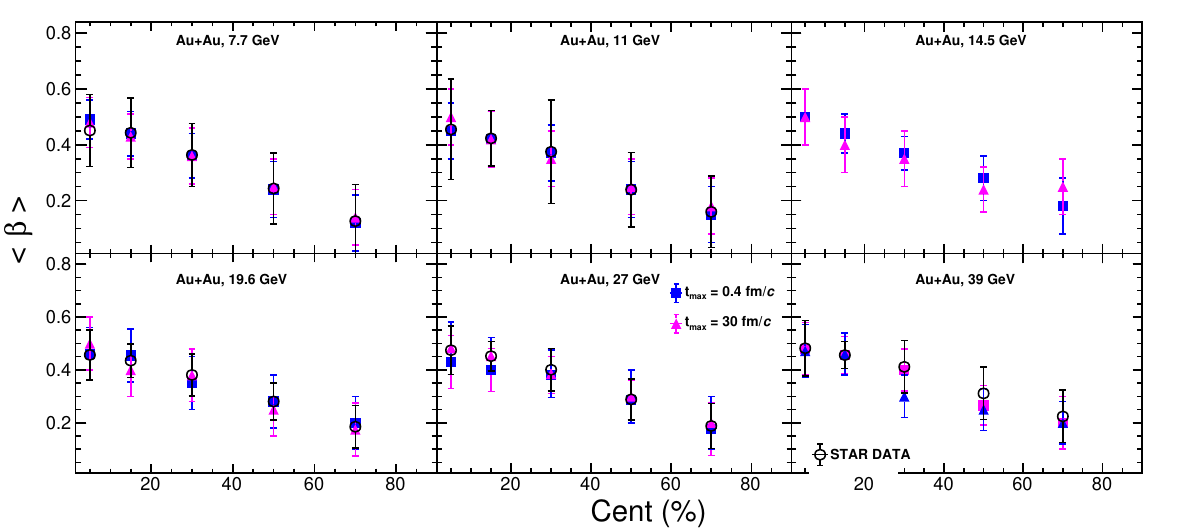}
    \caption{Transverse flow velocity, $<\beta_T>$ as a function of centrality in {\auau} collisions at {\sqrtsNN} = 7.7 - 39 GeV from AMPT-SM simulations for $t_{max}$ = 30 $f$m/$c$ and for $t_{max}$ = 0.4 $f$m/$c$. The STAR data is from Ref.~\cite{34}. No experimental data is available for {\sqrtsNN} = 14.5 GeV.}
    \label{fig4}
\end{figure*}
The spectra for energies, {\sqrtsNN} = 7.7, 11.5, 19.6, 27 and 39 GeV are also fitted with the Blast-wave model for both $t_{max}$ = 30 $f$m/$c$ and 0.4 $f$m/$c$. This procedure is used to extract $T_{kin}$ and $<\beta_T>$ and to study their dependence on energy and centrality for the two values of $t_{max}$. The results obtained are compiled in Table~\ref{table1} for $t_{max}$ = 0.4 $f$m/$c$ and in Table~\ref{table2} for $t_{max}$ = 30 $f$m/$c$.


\begin{figure*}[!ht]{\label{fig4a}} 
    \centering
    \includegraphics[width=\textwidth]{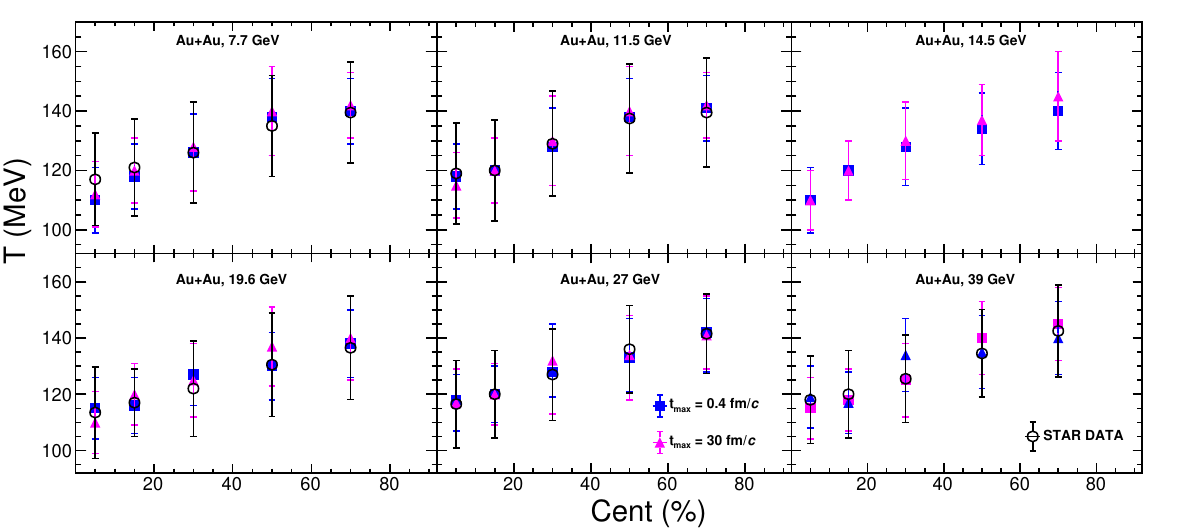}
    \caption{$T_{kin}$ as a function of centrality in {\auau} collisions at {\sqrtsNN} = 7.7 - 39 GeV from AMPT-SM simulations for $t_{max}$ = 30 $f$m/$c$ and for $t_{max}$ = 0.4 $f$m/$c$. The STAR data is from Ref.~\cite{34}. No experimental data is available for {\sqrtsNN} = 14.5 GeV.} 
    \label{fig4a}
\end{figure*}


The values of radial flow velocity, $<\beta_T>$ from Table~\ref{table1} and Table~\ref{table2} are shown graphically in fig.~\ref{fig4} for various energies and centrality classes. This figure additionally compares these values to the STAR experimental data~\cite{34}. Solid color symbols represent the AMPT-SM simulations while the open black symbols represent the STAR experimental data.  It is clear from the figure that there is good agreement between data and simulation for $<\beta_T>$ at all energies and centralities within statistical errors. All energies show a decreasing trend in $<\beta_T>$ with an increase in centrality, where large values of $<\beta_T>$ in most central collisions indicate a more rapid expansion. Figure~\ref{fig4} also shows similar trends for both values of $t_{max}$ which leads us to conclude that the $<\beta_T>$ does not depend on the values of $t_{max}$.  Even though experimental data is not available for {\sqrtsNN} = 14.5 GeV, we observe a similar trend in our AMPT-SM simulations for $<\beta_T>$ at this energy. Overall, there is a good agreement between experimental data and AMPT-SM simulations for both $t_{max}$ = 30 $f$m/$c$ and 0.4 $f$m/$c$.

\begin{figure*}[!ht]{\label{fig6}}
\centering
\includegraphics[width=0.6\textwidth]{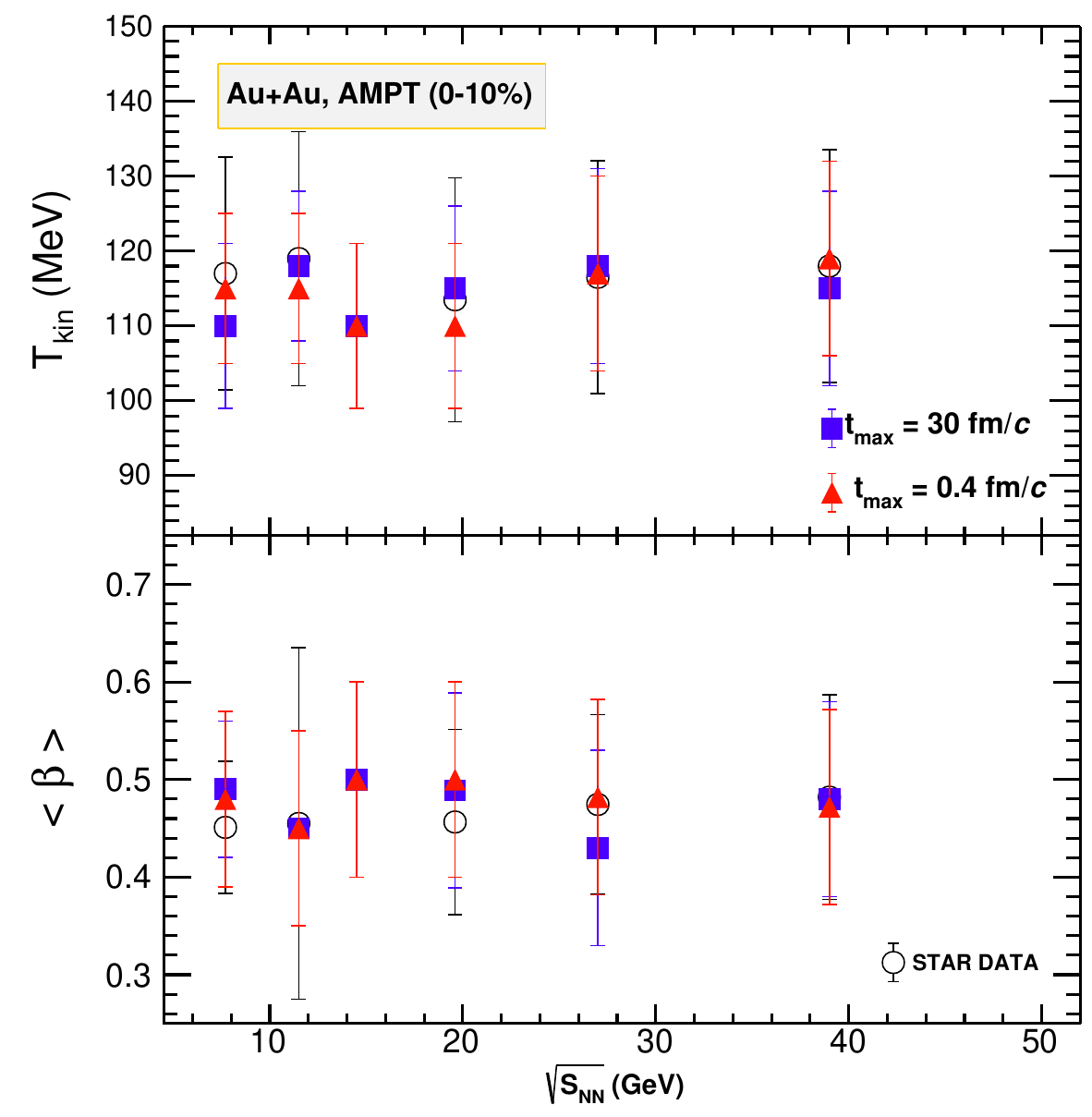}
\caption{Collision energy dependence of the extracted kinetic freeze-out temperature, $T_{kin}$ (upper panel) and transverse flow velocity, $<\beta_T>$ (lower panel) in {\auau} collisions at {\sqrtsNN} = 7.7 - 39 GeV from AMPT-SM model for 0-10\% centrality in blast wave fit to the {\ppt} spectra of identified hadrons.}
\label{fig6}
\end{figure*}


 The centrality dependence of the kinetic freeze-out temperature, $T_{kin}$ is shown in fig.~\ref{fig4a} as solid symbols, whereas the STAR experimental data from Ref.~\cite{34} is shown as open black symbols. There is a clear agreement between $T_{kin}$ obtained from AMPT-SM simulations and the experimental data within statistical errors. It is also observed that for AMPT-SM simulations $T_{kin}$ increases from central to peripheral collisions. This trend is as expected and is consistent with short-lived fireball in most peripheral collisions~\cite{34, 37}.  
 As mentioned above, no experimental results at {\sqrtsNN} = 14.5 GeV for {\auau} collisions are available, so a new measurement of $T_{kin}$ at this energy from AMPT-SM simulations are given in fig.~\ref{fig4a}. We observe that the $T_{kin}$  also increases from central to peripheral collisions.  By comparing two different values for the parameter $t_{max}$ in AMPT-SM simulations, we observe no significant difference in the value of $T_{kin}$. Overall, there is a good agreement between the AMPT-SM simulations and experimental data at all energies.

There is an additional interesting aspect, the centrality dependence of the fit parameters of figs.~\ref{fig4} and \ref{fig4a} that is worth discussing at this point. We fitted a wide range of centrality classes and observed that less radial flow is developed in most peripheral collisions. This means that there is lower freeze-out at higher temperatures, since there exists a strong anti-correlation between freeze-out temperature and flow~\cite{45}, which can also be seen by the slope of the {\ppt} spectra. This argument is consistent with the expectation that the fireball created just after the collision in most peripheral collisions, where a small number of participating nucleons ($N_{part}$) take part in the collision, do not have a longer lifetime and hence have less time to build the radial flow. However, in $pp$ collisions one would expect little to even no collectivity~\cite{44}. It is clear from the figs.~\ref{fig4} and ~\ref{fig4a} that in most peripheral {\auau} collisions the transition is steeper, because fewer $N_{part}$ take part in the collision and hence produce significant collectivity and a sizeable radial flow. This argument also holds true for the AMPT-SM simulations performed in the current study for both values of $t_{max}$.

Figure~\ref{fig6} shows the collision energy dependence of $T_{kin}$ in the upper panel and $<\beta_T>$ in the lower panel in {\auau} collisions at {\sqrtsNN} = 7.7 - 39 GeV for 0 - 10\% centrality from AMPT-SM simulations when compared to experimental data from STAR~\cite{34}. The APMT-SM simulation results presented here study the effect of hadronic cascade time to the extracted parameters for $t_{max}$ = 30 $f$m/$c$ and 0.4 $f$m/$c$. 
It is observed that for both values of $t_{max}$, $T_{kin}$ and $<\beta_T>$ at 0-10\% centrality show a weak collision energy dependence and is almost constant for {\sqrtsNN} = 7.7 - 39 GeV. Additionally, no significant impact is observed on the kinetic freeze-out parameters by changing the value of $t_{max}$ at these energies. However, it has been reported~\cite{Lokesh, 46} that $T_{kin}$ decreases towards higher energies. There is a good agreement with the experimental data and the AMPT-SM simulations for all energies. We also report the APMT-SM simulations at {\sqrtsNN} = 14.5 GeV, where no experimental data is yet available for comparison and we observe a similar trend here as is present at other energies.

\begin{figure*}[!ht]{\label{fig7}}
\centering

\includegraphics[width=\textwidth]{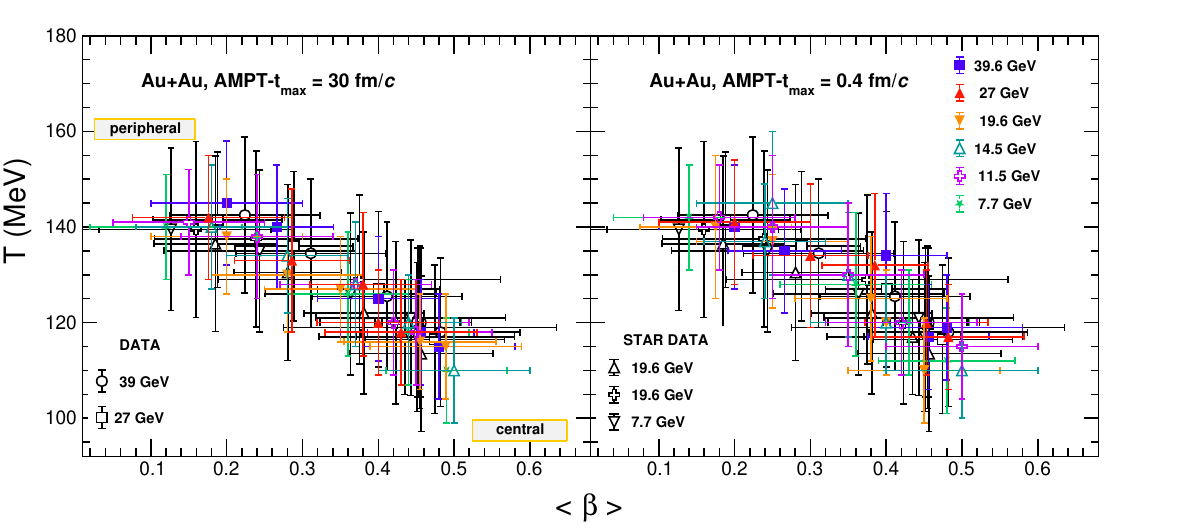}
\vspace{-1\baselineskip}
\caption{Variation of $T_{kin}$ with $<\beta_T>$ in {\auau} collisions at {\sqrtsNN} = 7.7 - 39 GeV from AMPT-SM model for $t_{max}$ = 30 $f$m/$c$ (left) and  $t_{max}$ = 0.4 $f$m/$c$ (left) for various centralities and energies. For a given energy the centrality increases from left to right. The experimental data is from Ref.~\cite{34}. No experimental data is available for {\sqrtsNN} = 14.5 GeV.} 
\label{fig7}
\end{figure*}


The variation of $T_{kin}$ with $<\beta_T>$ in {\auau} collisions at {\sqrtsNN} = 7.7 - 39 GeV from AMPT-SM model for different centralities and different values of $t_{max}$ are shown in fig.~\ref{fig7}. The colored symbols of different styles show the AMPT-SM simulations, while open black symbols represent experimental data from STAR~\cite{34}. For all energies, the figure shows that the centrality increases from left to right. The $<\beta_T>$ shows a decreasing trend from central to peripheral collisions indicating more rapid expansion in central collisions when compared to peripheral collisions for both values of $t_{max}$. On the other hand, $T_{kin}$ increases from central to peripheral collisions which is consistent with the expectations that due to fewer $N_{part}$, a fireball lives shorter in peripheral collisions~\cite{46}. Further, similar to the experimental results, we observe that the AMPT-SM simulations show a similar behaviour and the parameters show a strong anti-correlation, i.e., a higher $T_{kin}$ corresponds to a lower $<\beta_T>$  and vice versa. There is no significant effect observed by changing the $t_{max}$ value. Overall, the AMPT-SM simulations well describe the experimental data at all BES energies.  Again, we also report these parameters in {\auau} collisions at {\sqrtsNN} = 14.5 GeV from AMPT-SM simulations where experimental data is not yet published. Here we also observe similar trends as those observed at other energies.

\section{Conclusions}
\label{Conclusion}
In this study, we report the transverse momentum {\ppt} spectra of identified hadrons ($\pi^\pm$, $K^\pm$ and $p (\bar p)$) in {\auau} collisions at {\sqrtsNN} = 7.7 - 39 GeV from the improved version of AMPT-SM model with a different set of parameters. To study the effect of hadronic cascade time on the {\ppt} spectra and freeze-out parameters we chose two values of the parameter $t_{max}$, 30 $f$m/$c$ and 0.4 $f$m/$c$ and the {\ppt} spectra presented here is for 0-10\%, 10-20\%, 20-40\%, 40-60\% and 60-80\% centrality classes. The AMPT-SM simulation well describes the identified hadrons {\ppt} spectra and we observe no significant difference between the two $t_{max}$ values. 

We also studied the kinetic freeze-out parameters, $T_{kin}$ and $<\beta_T>$ extracted from the blast wave fit to AMPT-SM simulations for both values of $t_{max}$ and compared it with available experimental data. We observed that $T_{kin}$ shows an increasing trend from central to peripheral collisions, which indicates a long-lived fireball for central collisions when compared to peripheral collisions. This is due to a large number of participating nucleons ($N_{part}$) contributing in the collisions in central collisions. On the other hand, $<\beta_T>$ shows a decreasing trend from central to peripheral collisions, which indicates a rapid expansion of the fireball in central collisions when compared to the peripheral collisions. These parameters show a weak dependence on the collision energy for both $t_{max}$ values in AMPT-SM simulations, which is consistent with experimental data.  The $<\beta_T>$ and $T_{kin}$ show a strong anti-correlation for both values of $t_{max}$, i.e. lower value of $<\beta_T>$ corresponds to higher value of $T_{kin}$, similar to the experimental observations.  

Overall, the AMPT-SM model with the current set of parameters reproduce the identified hadrons {\ppt} spectra successfully. The {\ppt} spectra is not affected by change in hadronic cascade parameter ($t_{max}$). The blast wave model reasonably fits the AMPT-SM simulations  and the extracted kinetic freeze-out parameters are in good agreement with the data obtained from the STAR experiment. Further, no significant difference is observed in the values of kinetic freeze-out temperature, $T_{kin}$ and the transverse flow velocity, $<\beta_T>$ by changing $t_{max}$. With our set of parameters we do not see the discrepancy between AMPT-SM simulations and experimental data reported in earlier studies. 
 Hence, we conclude that the set of parameters  used for this study better describes the {\ppt} spectra and the kinetic freeze-out parameters measured by the STAR experiment. In summary, our study provides a reference for {\auau} system scan and kinetic freeze-out properties of hot and dense QCD matter created in heavy-ion collisions at RHIC-BES energies.

\begin{table*}[!ht]
\scriptsize{
\caption{Extracted kinetic freeze-out parameters in {\auau} collisions at {\sqrtsNN} = 7.7 - 39 GeV from AMPT-SM simulations for $t_{max}$ = 0.4 $fm/c$ quoted with errors.}
\vspace{-.50cm}
\begin{center}
\begin{tabular}{p{2cm}p{2cm}p{2cm}p{2cm}p{2cm}p{2cm}}\\ \hline\hline
     Collision & Centrality &   $<\beta> c$ &    $T$ (MeV) & Probability & $\chi^2$/ dof  \\\hline
    Au-Au 7.7 GeV   & $0-10\%$ & $0.48\pm0.09$   & $112\pm11$        &$0.60$ & 0.600\\
            & $10-20\%$ &$0.43\pm0.08$   & $120\pm11$        &$ 0.87$  & 0.360  \\
            & $20-40\%$ &$0.36\pm0.10$   & $128\pm15$        &$0.52$ & 0.820\\
            & $40-60\%$ &$0.25\pm0.10$   & $140\pm15$        &$ 0.77$   & 0.700 \\
            & $60-80\%$     &$0.14\pm0.10$   & $142\pm15$        &$ 0.33$  & 1.200\\
            \hline
    Au-Au 11.5 GeV   & $0-10\%$ & $0.50\pm0.10$   & $115\pm11$        &$0.60$ & 0.660\\
            & $10-20\%$ &$0.42\pm0.10$   & $120\pm11$        &$ 0.87$  & 0.400  \\
            & $20-40\%$ &$0.35\pm0.10$   & $130\pm15$        &$0.52$ & 0.840\\
            & $40-60\%$ &$0.25\pm0.10$   & $140\pm15$        &$ 0.77$   & 0.500 \\
            & $60-80\%$     &$0.18\pm0.10$   & $142\pm10$        &$ 0.33$  & 1.142\\
  \hline
 Au-Au 14.5 GeV   & $0-10\%$ & $0.50\pm0.10$   & $110\pm10$        &$0.93$ & 0.250\\
            & $10-20\%$ &$0.40\pm0.08$   & $120\pm11$        &$ 0.63$  & 0.680 \\
            & $20-40\%$ &$0.35\pm0.07$   & $130\pm13$        &$ 0.65$ & 0.660\\
            & $40-60\%$ &$0.24\pm0.08$   & $137\pm12$        &$ 0.47$   & 0.974 \\
            & $60-80\%$     &$0.20\pm0.10$   & $145\pm10$        &$ 0.40$  & 1.190\\
  \hline
   Au-Au 19.6 GeV   & $0--10\%$ & $0.50\pm0.10$   & $110\pm11$        &$0.71$ & 0.582\\
            & $10-20\%$ &$0.40\pm0.07$   & $120\pm11$   &$ 0.48$  & 0.930  \\
            & $20-40\%$ &$0.38\pm0.07$   & $125\pm13$        &$0.70$ & 0.600\\
            & $40-60\%$ &$0.25\pm0.08$   & $137\pm15$   &$ 0.85$   & 0.400 \\
            & $60-80\%$     &$0.17\pm0.10$   & $142\pm10$        &$ 0.40$  & 1.140\\
  \hline
  Au-Au 27 GeV   & $0-10\%$ & $0.48\pm0.10$   & $117\pm11$        &$0.67$ & 0.632\\
            & $10-20\%$ &$0.45\pm0.07$   & $120\pm11$        &$ 0.58$  & 0.760  \\
            & $20-40\%$ &$0.38\pm0.09$   & $132\pm15$        &$0.97$ & 0.200\\
            & $40-60\%$ &$0.30\pm0.10$   & $134\pm15$        &$ 0.73$   & 0.941 \\
            & $60-80\%$     &$0.20\pm0.10$   & $141\pm10$        &$ 0.77$  & 0.510\\
  \hline
   Au-Au 39 GeV   & $0-10\%$ & $0.47\pm0.10$   & $119\pm11$        &$0.60$ & 0.730\\
            & $10-20\%$ &$0.46\pm0.08$   & $117\pm11$        &$ 0.33$  & 1.162  \\
            & $20-40\%$ &$0.30\pm0.08$   & $134\pm13$        &$0.30$ & 1.140\\
            & $40-60\%$ &$0.25\pm0.08$   & $135\pm13$        &$ 0.80$   & 0.480 \\
            & $60-80\%$     &$0.20\pm0.08$   & $140\pm13$        &$ 0.10$  & 2.410\\
    \hline
 \hline
\end{tabular}
\label{table1}
\end{center}} 
\end{table*}

\begin{table*}[!ht]
{\scriptsize 
\caption{Extracted kinetic freeze-out parameters in {\auau} collisions at {\sqrtsNN} = 7.7 - 39 GeV from AMPT-SM simulations for $t_{max}$ = 30 $fm/c$ quoted with errors.}
\vspace{-.50cm}
\begin{center}
\begin{tabular}{p{2cm}p{2cm}p{2cm}p{2cm}p{2cm}p{2cm}}\\ \hline\hline
     Collision & Centrality &   $<\beta> c$ &    $T$ (MeV) & Probability & $\chi^2$/ dof  \\\hline
   Au-Au 7.7 GeV   & $0-10\%$ & $0.49\pm0.07$   & $110\pm11$        &$0.40$ & 0.446\\ & $10-20\%$ &$0.44\pm0.08$   & $118\pm11$        &$ 0.92$  & 0.592  \\ & $20-40\%$ &$0.36\pm0.08$   & $126\pm13$        &$0.50$ & 0.550 \\ & $40-60\%$ &$0.24\pm0.10$   & $138\pm13$        &$ 0.71$   & 1.208 \\ & $60-80\%$     &$0.12\pm0.10$   & $140\pm11$        &$ 0.90$  & 1.160\\
    \hline
   Au-Au 11.5 GeV   & $0-10\%$ & $0.45\pm0.10$   & $118\pm11$        &$0.40$ & 1.052\\
            & $10-20\%$ &$0.42\pm0.10$   & $120\pm11$        &$ 0.92$  & 0.290  \\
            & $20-40\%$ &$0.37\pm0.10$   & $128\pm13$        &$0.50$ & 0.864\\
            & $40-60\%$ &$0.24\pm0.10$   & $138\pm13$        &$ 0.71$   & 0.580 \\
            & $60-80\%$     &$0.15\pm0.10$   & $142\pm10$        &$ 0.90$  & 0.350\\
  \hline
 Au-Au 14.5 GeV   & $0-10\%$ & $0.50\pm0.10$   & $110\pm10$        &$0.45$ & 1.008\\
            & $10-20\%$ &$0.44\pm0.07$   & $120\pm10$        &$ 0.70$  & 0.596  \\
            & $20-40\%$ &$0.37\pm0.06$   & $128\pm13$        &$ 0.65$ & 0.670\\
            & $40-60\%$ &$0.28\pm0.08$   & $134\pm12$        &$ 0.48$   & 0.902 \\
            & $60-80\%$     &$0.18\pm0.10$   & $140\pm13$        &$ 0.95$  & 0.180\\
  \hline
   Au-Au 19.6 GeV   & $0-10\%$ & $0.48\pm0.11$   & $115\pm11$        &$0.51$ & 1.008\\
            & $10-20\%$ &$0.45\pm0.07$   & $116\pm10$   &$ 0.60$  & 0.754  \\
            & $20-40\%$ &$0.35\pm0.07$   & $127\pm11$        &$0.59$ & 0.750\\
            & $40-60\%$ &$0.28\pm0.08$   & $130\pm12$   &$ 0.70$   & 0.760 \\
            & $60-80\%$     &$0.20\pm0.10$   & $138\pm12$        &$ 0.91$  & 0.328\\
  \hline
  Au-Au 27 GeV   & $0-10\%$ & $0.43\pm0.10$   & $118\pm11$        &$0.70$ & 0.600\\
            & $10-20\%$ &$0.40\pm0.07$   & $120\pm11$        &$ 0.70$  & 0.600  \\
            & $20-40\%$ &$0.38\pm0.07$   & $128\pm13$        &$0.50$ & 0.860\\
            & $40-60\%$ &$0.28\pm0.10$   & $133\pm13$        &$ 0.70$   & 0.560 \\
            & $60-80\%$     &$0.17\pm0.10$   & $142\pm10$        &$ 0.83$  & 0.416\\
  \hline
   Au-Au 39 GeV   & $0-10\%$ & $0.48\pm0.10$   & $115\pm11$        &$0.80$ & 0.468\\
            & $10-20\%$ &$0.45\pm0.07$   & $118\pm11$        &$ 0.80$  & 0.440  \\
            & $20-40\%$ &$0.30\pm0.07$   & $134\pm13$        &$0.30$ & 0.860\\
            & $40-60\%$ &$0.40\pm0.10$   & $125\pm13$        &$ 0.57$   & 0.820 \\
            & $60-80\%$     &$0.20\pm0.09$   & $145\pm13$        &$ 0.20$  & 1.700\\
  \hline
\end{tabular}
\label{table2}
\end{center}}  
\end{table*}

\bibliography{bib.bib}
\bibliographystyle{unsrt}
\end{document}